# D-separation for applied researchers: understanding how to interpret directed acyclic graphs

Fernando Pires Hartwig[1,2]
Timothy Feeney[3]
Neil Davies[4,5,6]

[1] Postgraduate Program in Epidemiology, Federal University of Pelotas, Pelotas, Brazil.
[2] Medical Research Council Integrative Epidemiology Unit at the University of Bristol, BS8 2BN, United Kingdom.
[3] Gillings School of Global Public Health, Department of Epidemiology, University of North Carolina at Chapel Hill, Chapel Hill, NC USA
[4] Division of Psychiatry, University College London, Maple House, 149 Tottenham Court Rd, London W1T 7NF
[5] Department of Statistical Sciences, University College London, London WC1E 6BT, UK
[6] K.G. Jebsen Center for Genetic Epidemiology, Department of Public Health and Nursing, Norwegian University of Science and Technology, Norway.

Correspondence to: Fernando Pires Hartwig, fernandophartwig@gmail.com

Word count (main text): 2937 words. References: 8.

## Contributors and sources
FPH, TF and NMD conceived the paper. FPH wrote the first draft. FPH, TF and NMD revised it.

## Conflicts of Interest
We have read and understood BMJ policy on declaration of interests and have the following interests to declare: TF is a research editor at the BMJ and a clinical editor at BMJ Medicine. NMD receives funds from the BMJ and Addiction for editorial work.

## Funding statement
NMD is supported via a Norwegian Research Council Grant (295989) and the UCL Division of Psychiatry (https://www.ucl.ac.uk/psychiatry/division-psychiatry). FPH is supported by the Brazilian National Council for Scientific and Technological Development (303880/2023-6).

**Standfirst**

*The assumed causal relationships depicted in a DAG are interpreted using a set of rules called D-separation rules. Although these rules can be implemented automatically using standard software, at least a basic understanding of their principles is useful for properly using and interpreting DAGs in practice.*

**Introduction**

Causal Directed acyclic graphs (hereafter DAGs) are widely used to illustrate assumptions about causal relationships between variables - see Feeney and colleagues (2024) for an introduction to DAGs and terminology that will be used throughout this text[1]. DAGs are useful not only for transparency but also for guiding data analysis in accordance with the assumptions one is willing to make. For example, DAGs allow for the identification of covariates that should be adjusted to prevent confounding or selection bias; DAGs also aid in identifying covariates that should not be adjusted because doing so would introduce bias[2,3]. All these applications rely on applying a set of rules - called d-separation rules - to interpret the causal structure (and its implications) depicted in the DAG. Here, we explain what d-separation rules are in an accessible way to enable applied researchers to have a more complete understanding of how DAGs are used and interpreted.

In a DAG, exposure (E) and outcome (O) can be linked to one another through one or more causal paths ('edges') and one or more biasing (or non-causal) paths. Causal paths are paths of form E → … → O - i.e., the path points from E into O, and all edges are directed similarly. These paths indicate all direct and indirect effects. A biasing path can be understood as any path between exposure and outcome that is not causal.

A path may be open or blocked: intuitively, an open path between two variables ('nodes') is a source of association between them. Conversely, a blocked path cannot explain an association between them. Causal effect estimation requires that all causal paths of interest are open while all biasing paths and non-target causal paths are blocked. This way, the only source of association (if any) between exposure and outcome will be the causal effect of interest. This can be achieved by applying the d-separation rules to the DAG at hand.

**D-connection and d-separation**

Two variables are considered d-separated when they have no open path. Otherwise (i.e., when there is at least one open path between them), they are d-connected[2,3]. We now explain the concepts of d-separation and d-connection in relation to an exposure-outcome association.

First, consider the case where no variable is being adjusted explicitly (e.g., due to analytical decisions) or implicitly (e.g., due to missing data, which results in adjustment by restricting the analysis to those with non-missing data). In this case, a path between exposure and outcome is open if the path contains no collider. Otherwise (i.e., if the path contains one or more colliders), the path is closed. So, exposure and outcome could be d-connected via paths such as E→O (a direct causal path), E→M→O (an indirect causal path) and E←C→O (a biasing path due to confounding). For example, in Figure 1A, there is no direct causal path from "nutrition" to "plays basketball". However, this does not imply they have no causal relationship. Indeed, in this example, there is one path between them: the path Nutrition→Height→Plays basketball. Since this path contains no collider, it is open. More specifically, it is an indirect causal path, because the effect of nutrition is mediated by height.

It is rather intuitive that paths of this form correspond to sources of association between E and O: causal paths (either direct or indirect) imply that changing E would change O, while a confounding path implies that changing C would change both E and O, thus causing them to correlate. Paths of the form E→M→O and E←C→O would be intuitively blocked by adjustment/conditioning/stratification for M (because the causal chain from E to O is broken) or C (because, within strata of C, C is not a source of covariation between E and O), respectively[2,3].

**D-separation and colliders**

A path between two variables containing one or more colliders not being adjusted for (adjusting for colliders is discussed below) is blocked[2,3]. The fact that one or more (unadjusted) colliders block the path may seem counterintuitive at first, but follows from the fact that two causes affecting the same consequence does not imply the causes are themselves related. If this were not the case, it would be impossible for independent causes of the same effect to exist. Consider a simple example: both biological sex and nutrition in childhood and adolescence influence height. However, this does not imply that sex and nutrition are associated with one another: they could simply be independent causes of height. Of course, the two causes can be correlated; the point is that if they are related, it is not because they have a common effect.

Adjustment for a collider opens the path at the collider. In other words, a path of the form E→K←O would be opened by adjusting for K. Note the path E→K←O, opened by adjusting for K, is a biasing path, so opening it would bias the association between E and O. This is known as collider bias[4]. Consider the example illustrated in Figure 1A, where we assume height is a consequence of both sex at birth and nutritional quality. Suppose that sex and nutrition are independent in the overall population. If one considers only tall people, then knowing that the individual was a female leads to a greater probability that their nutrition was good than if they were male. This is because it is rarer for an undernourished female than an undernourished male to be tall, thus leading to an association between the two causes. Moreover, conditioning on a consequence of a collider, in this example, whether a person plays basketball (influenced by height), can also open a biasing path between sex at birth and nutrition, similarly to how conditioning directly on the collider can open the same path. This is because a consequence of a collider is itself a collider - in this example, playing basketball is a common effect of sex and nutrition. Thus, conditioning on the consequence of a collider can still lead to collider bias (see **Box 2** for another example).

**Applying d-separation rules to a DAG: a step-by-step example**

Figure 1B illustrates a hypothetical DAG, which we will use to illustrate the application of d-separation rules. In this DAG, U1, U2 and U3 are unmeasured variables, C1 and C2 are measured covariates, M1 and M2 are mediators of the effect of the exposure (E) on the outcome (O), and S represents eligibility criteria for the study. Importantly, S is being adjusted for because any analysis will necessarily be restricted to the stratum of S corresponding to being eligible. For example, if S=1 denotes eligible individuals and S=0 denotes non-eligible individuals, the analysis is restricted to the stratum S=1.

We discuss two research questions differing with regards to the causal effect one is interested in estimating: 1) the total effect of E on O (i.e., through both M1 and M2); and 2) the indirect effect of E on O through M1 only. In both cases, one needs to use d-separation rules to block all paths other than the target causal path(s).

For research question 1, one needs to block all non-causal paths. This way, the only possible sources of association between E and O are the two causal paths E → M1 → O and E → M2 → O. More specifically, in this example, one needs to block the non-causal paths E ← C1 → S ← U1 → O and E ← C1 → S ← U1 → M1 → O. Notice these paths are open because the single collider they contain (S) is being adjusted for. Both paths could, in principle, be blocked by adjusting for C1 and/or U1. However, since U1 is unmeasured, the only possibility would be C1. Therefore, C1 would be a valid adjustment set for estimating the total effect of E on O. Of note, the non-causal path E ← U2 → M2 ← U3 → C2 → O is blocked because it contains a collider (M2) that is not being adjusted for.

Now, consider research question 2. In this case, one needs to block all other causal paths between E and O (because those would be non-target causal paths) and all non-causal paths. This way, the only possible source of association between E and O is the target causal path.

More specifically, in this example, one needs to block: i) E ← C1 → S ← U1 → O and E ← C1 → S ← U1 →M1 → O and ii) the causal path E → M2 → O. Paths i) are the same paths mentioned above, which can be blocked by adjusting for C1 since U1 is unmeasured. Path ii) can only be blocked by conditioning on M2. But doing so would open the non-causal path E ← U2 → M2 ← U3 → C2 → O (notice M2 is now being adjusted for to block path ii), which can be blocked by adjusting for C2 (notice U2 and U3 are unmeasured). Therefore, by adjusting for M2, C1 and C2, the only open path remaining between E and O is the path E → M1 → O, which is the causal path of interest. The set {C1, C2, M2} is a valid adjustment set for estimating the indirect effect of E on O through M1.

This example illustrates that, when selecting adjustment variables to block a path, it is important to check if adjusting for the variable induces bias by opening a non-causal path via collider bias. If this happens, one alternative is to select a different, non-collider variable for adjustment. However, sometimes, it is impossible to not adjust for a collider. In the example above, it was necessary to adjust for M2 because the research question involved only the causal path through M1. In such cases, it is necessary to find adjustment variables that block such open non-causal paths as a result of the variables being adjusted for (in the example, C2).

**Other applications of d-separation rules**

In the examples above, we described the importance of d-separation rules for using DAGs for covariate selection, which is one of the most frequent applications of DAGs for causal effect estimation. However, it is important to realise that d-separation rules underlie any application of DAGs for causal inference, since these are the rules used to interpret the causal structure of the DAG. To illustrate this, we now briefly mention two additional applications of d-separation rules when interpreting DAGs.

One of them relates to generalizability/transportability. To illustrate this, we refer to Figure 1B again, which contains a node S for selection into the study, which is influenced by C1 and U1. Now, suppose instead that S denotes "having complete data for E, D, C1 and C2". In a cohort study, this would be the individuals who were not lost to follow-up and were eligible to all relevant examinations. Above, we described that it was possible to estimate the causal effect of E on O by adjusting for measured covariates. As mentioned above, the analysis is necessarily restricted to the stratum S=1. That is, our results only apply to those with non-missing data. However, we are interested in estimating the causal effect for the whole cohort, not only in the subset of individuals with non-missing data. DAGs can be used to assess if the results in the stratum S=1 can be extrapolated to the stratum S=0. Briefly, this is not in general possible when S and O are d-connected, because in this case S can be a (proxy) modifier of the effect of E on O, which implies the possibility that the effect of E on O varies between strata of S (see Hernán (2017) for more details)[5]. In Figure 1B, S and O are d-connected through the path S ← U1 → O. Since U1 is unmeasured, this path cannot be blocked. Had U1 been measured, it would have been possible to block this path by weighting or standardisation[2].

The second additional application of d-separation rules to DAGs we will mention relates to instrumental variable (IV) analysis (see Hernán and Robins 2006 and Walker and Colleagues (2024) for introductions to instrumental variables)[6,7]. A valid IV for the effect of E on O is a variable that satisfies the following conditions: (i) relevance: the IV and E and robustly statistically dependent; (ii) independence: the IV and O have no common causes such that the backdoor path between IV and O does not contain the causal path from E to O; (iii) exclusion restriction: there is no causal path from the IV to O that is not mediated by E. Paths (ii) and (iii) can be combined as follows: there is no open path between the IV and O that does not contain the causal path from E to O. Of note, condition (i) requires that the IV and E are d-connected, and can be empirically verified in the data. So, it is possible to apply d-separation rules to a DAG to check if the DAG suggests that one or more variables are valid IVs (possibly after adjusting for one or more variables) by checking if the candidate IV and O are: (1) d-connected in a DAG containing any causal path from E to O; and (2) d-separated in a DAG where, only for all causal paths from E to O, the arrows originating from E are removed. If both conditions are satisfied, then Z can only be statistically associated with O (apart from chance, of course) if E has a causal effect on O. As will be illustrated below, this is true even if, for example, there is unmeasured confounding between E and O. So, if the candidate IV is truly valid, then it can be used to study the causal effect of E on O even if methods based on covariate adjustment fail.

Let us apply D-separation rules to the DAGs in Figure 2 to check if Z is a valid IV in the stratum S=1. Notice we have the same DAG twice, except that the causal path from E to O that exists in panel A does not exist in panel B because the latter does not contain an arrow from E to M. So, if Z is a valid IV, it must be d-connected with O in panel A, but d-separated in panel B. Notice that Z is d-connected with O in both panels, so Z is not a valid IV. In panel B, Z and O are d-connected through two paths: Z → C2 → M → O and Z ← U1 → E → S ← C1 → O (open because we are restricting to S=1). However, both paths contain measured non-colliders: C1 and C2 (M cannot be chosen instead of C2 because it mediates the effect of E on O). So, upon adjustment for these variables, Z is a valid IV since it becomes d-separated with O in panel B but remains d-connected with O in panel A through the path Z ← U1 → E → M → O (which is a backdoor path between Z and O, but contains the causal path from E to O, so this path is only open because E causes O in this DAG). Of note, had we removed the causal path from E to O in panel B by removing the arrow from M to O instead of the arrow from E to M, the path Z → C2 → M → O would have also been eliminated from the DAG, and we would not have noticed that the latter also needs to be blocked by adjusting for C2.

**Final remarks**

Given the discussion above, a simple way to summarise the d-separation rules is the following: a path between E and O is open when there is no unadjusted collider (since one unadjusted collier would suffice to block the path) and no adjusted non-collider (since one adjusted non-collider would suffice to block the path). Otherwise, the path is blocked.

It is important to emphasise that, as any causal inference enterprise, the inferences' validity depends on the assumptions' validity[2]. Whether the adjustment sets selected in the examples above are valid (i.e., they block all non-causal and non-target causal paths while leaving target causal paths open) depends on the validity of the assumptions encoded in the DAG. The procedures we described (as well as any other conclusion obtained by applying d-separation rules to a DAG) assume the DAG is true. Therefore, the most important (and often most difficult) step is drawing the DAG, which requires justification based on expert knowledge[1].

Some limitations of D-separation rules should also be acknowledged. The most important one is that they assume the DAG is correct - i.e., they simply interpret the DAG drawn. The correctness of any DAG representing an empirical situation is virtually always debatable, and so will be the conclusions resulting from applying d-separation rules to a DAG. Moreover, it is assumed that the variables were perfectly measured (it is possible to incorporate measurement error nodes in a DAG, but this is rarely performed) and their statistical relationships were correctly modelled[1]. Violations of either measurement error or model misspecification can lead to bias, even if the DAG suggests that measured covariates are sufficient for bias elimination. Considering these limitations when interpreting the results obtained by applying D-separation rules to a DAG is important.

In practice, conclusions about the status of a given path (causal or non-causal; open or blocked) and covariate selection strategies are determined by computer software since they follow algorithmically from a given DAG[8]. Nevertheless, at least a basic understanding of the principles underlying how DAGs are interpreted is useful for properly applying these tools.

**Figure 1. Directed acyclic graphs (DAGs) illustrating the use of D-separation rules to select variables for statistical adjustment.**

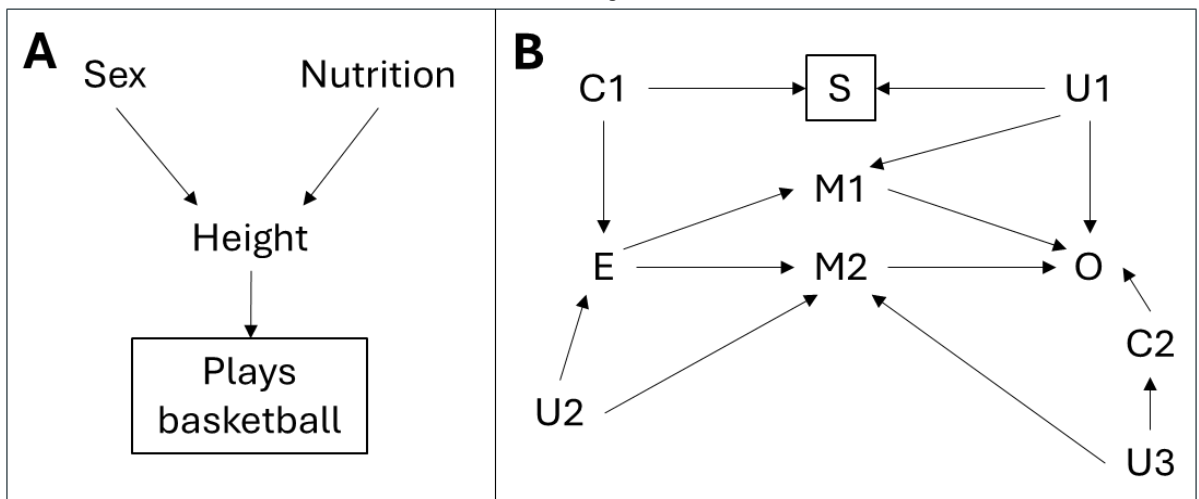

A) Hypothetical DAG where sex and nutrition are assumed to be independent causes of height. The box around "plays basketball" indicates that this variable is being adjusted. In our example, we consider restricting to those who play basketball. Since playing basketball is a consequence of the collider height, sex and nutrition become d-connected upon adjustment, even though they would be d-separated had no adjustment been done. Intuitively, this happens because basketball players are, on average, taller than people who do not. So, restricting to this subset induces an association between sex and nutrition similar to restricting to tall people (described in the main text). B) Hypothetical DAG. U1-3 are unmeasured variables, C1-2 are measured covariates, M1 and M2 are mediators of the effect of the exposure (E) on the outcome (O), and S represents eligibility criteria for the study (notice that S is being adjusted for because any analysis will be restricted to the stratum of S corresponding to being eligible).

**Figure 2. Directed acyclic graphs (DAGs) illustrating the use of D-separation rules in relation to instrumental variable (IV) analysis.**

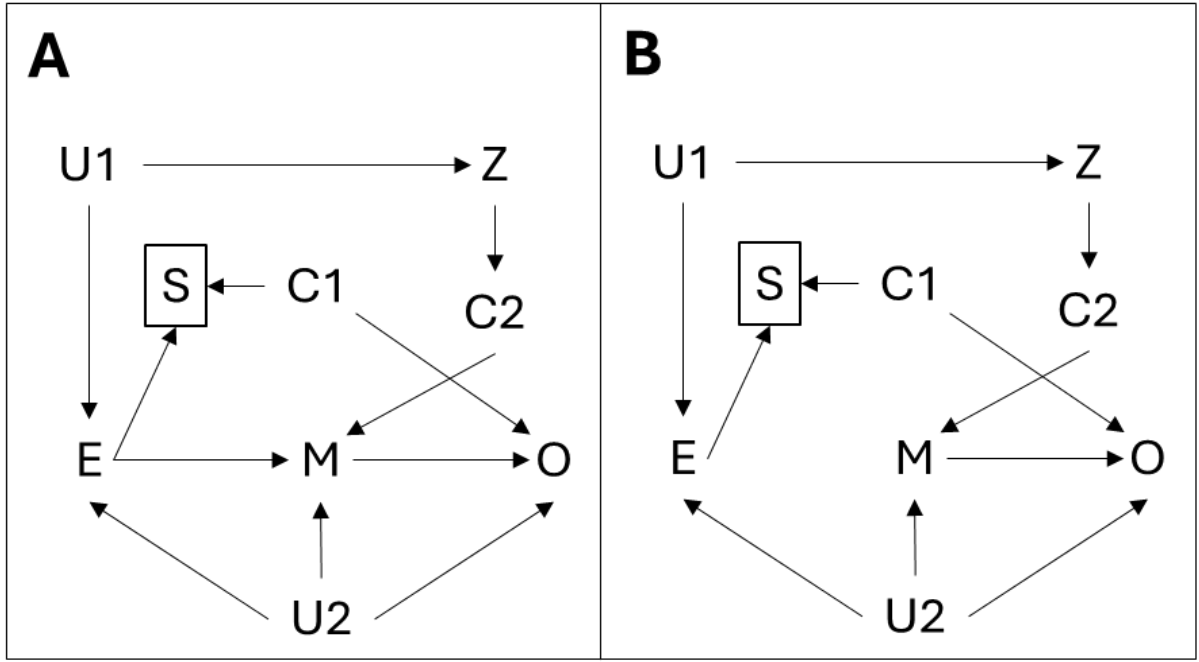

A) Hypothetical DAG. Z is a candidate IV, U-2 are unmeasured variables, C1-2 are measured covariates, M is a mediator of the effect of the exposure (E) on the outcome (O), and S represents eligibility criteria for the study (notice that S is being adjusted for because any analysis will be restricted to the stratum of S corresponding to being eligible). B) Same as A, except the arrow from E to M is missing.

**Box 1: Key messages**

- D-separation rules are the main principle underlying the interpretation of directed acyclic graphs (DAGs). They dictate whether or not there is (according to the DAG) bias and how to adjust for it.

- A causal path is directed from exposure to outcome, with all arrows in between being directed similarly. Multiple causal paths may correspond to direct and indirect effects. Any other path between exposure and outcome is non-causal.

- According to D-separation rules, a path between exposure and outcome is open (i.e., it is a source of association between them) if all colliders (i.e., common effects) in the path (if any) are being adjusted for (by stratification, restriction etc.) and all non-colliders are not being adjusted for. Otherwise, the path is blocked.

- One way to use DAGs for causal inference analysis can be briefly summarised as follows: by applying D-separation rules to the DAG, one ensures that the causal path of interest is open (i.e., it is a possible source of association between exposure and outcome) while all other paths are blocked. Whether or not this is valid depends on the validity of the assumptions encoded in the DAG.

**Box 2: Colliders, d-separation and coin flipping.**

Suppose we independently flip two fair coins and add the number of heads (denoted by H). Let C1 and C2, respectively, denote the outcome of coins 1 and 2. C1=0 corresponds to tails, and C1=1 corresponds to heads (similarly for C2). Then, H=C1+C2. H is a collider between C1 and C2. Suppose we repeat this experiment independently many times and write down the results.

By the nature of the experiment, C1 and C2 are independent. The rule that postulates that the presence of a collider in a path blocks the path simply allows such independence to exist - i.e., it simply states that causes of a common effect do not have to be correlated.

Suppose we restrict our attention to all scenarios where H=1 (i.e., we condition on H by restriction). In this subset of the experiments, if C1=1, then C2=0 (for their sum to be 0). Similarly, if C1=0, then C2=1. Therefore, in the subset H=1, C1 and C2 become associated even though they are independent before conditioning on H. In this example, conditioning on the collider H induced a negative association between C1 and C2.

Now, define a new variable Z as follows: Z=0 if H≤1, Z=1 if H=2. Clearly, Z is a consequence of the collider H. Suppose we restrict our attention to all scenarios where Z=0. In this subset, if C1=1, then C2=0; and if C2=1, then C1=0. Therefore, in the subset Z=1, C1 and C2 become associated even though they are independent before conditioning on Z. In this example, conditioning on Z - a consequence of the collider H - induced a negative association between C1 and C2.

This example illustrates why conditioning on a collider's consequence opens the path between its causes: it is because the collider's consequence is itself a collider. To see why this is the case in this example, note that Z could also have equivalently been defined as follows: Z=1 if C1=1 and C2=1; Z=0 otherwise. That is, Z is also a common effect of C1 and C2.